\documentclass[conference]{IEEEtran}

\usepackage{fancyhdr, etoolbox}
\fancypagestyle{firstpage}{%
    \fancyhead[L]{}
    \fancyhead[C]{The 21st IEEE International Conference on Machine Learning and Applications (ICMLA'22)}
    \fancyhead[R]{}
}

\usepackage[numbers, sort]{natbib}
\usepackage{amsmath,amssymb,amsfonts}
\usepackage{hyperref}
\usepackage{fancybox}
\usepackage{graphicx}
\usepackage{float}
\usepackage{url} 
\usepackage{adjustbox}
\usepackage{multirow}
\usepackage{xspace}
\usepackage{comment}
\usepackage{xcolor}
\usepackage{balance}
\usepackage{algorithmic}
\usepackage{algorithm}
\usepackage{cleveref} 
\usepackage{listings}
\usepackage[normalem]{ulem}
\usepackage{array}
\usepackage{pbox}
\usepackage{enumitem}

\usepackage[labelfont=bf]{caption}
\usepackage[labelfont=bf]{subcaption}
\newcolumntype{H}{>{\setbox0=\hbox\bgroup}c<{\egroup}@{}}

\newcommand{\Part}[1]{\noindent\textbf{#1}}

\newcommand{\Space}[1]{}

\newcommand{\eg}{\textit{e.g.}\xspace}
\newcommand{\ie}{\textit{i.e.}\xspace}

\cornersize{.2}
\newcounter{observation}
\newcommand{\observation}[1]{\refstepcounter{observation}
        \begin{center}
        \vspace{3pt}
        \Ovalbox{
        \begin{minipage}{0.9\columnwidth}
            \textbf{Observation \arabic{observation}:} #1
        \end{minipage}
        }
        \end{center}
}

\newcommand{\mnp}{\textsc{MethodPrediction}\xspace}
\newcommand{\mnps}{\textsc{Method Name Prediction}\xspace}
\newcommand{\cc}{\textsc{CodeClassification}\xspace}
\newcommand{\ccs}{\textsc{Code Classification}\xspace}

\newcommand{\JS}{\textsc{Java-Small}\xspace}

\newcommand{\JL}{\textsc{Java-Large}\xspace}

\newcommand{\SA}{\textsc{Java-Sort}\xspace}

\newcommand{\ctv}{\textsc{Code2Vec}\xspace}
\newcommand{\cts}{\textsc{Code2Seq}\xspace}

\newcommand{\ggnn}{\textsc{GGNN}\xspace}

\newcommand{\ftv}{\textsc{func2vec}\xspace}

\newcommand{\original}{\textsc{Original}\xspace}
\newcommand{\refactor}{\textsc{Reformatted}\xspace}
\newcommand{\xalpha}{\textsc{Redacted}\xspace}

\newcommand{\ttoken}{\textsc{Token}\xspace}
\newcommand{\tpath}{\textsc{Path}\xspace}

\newcommand{\bilstm}{\textsc{BiLSTM}\xspace}

\newcommand{\alexNet}{\textsc{AlexNet}\xspace}
\newcommand{\resNet}{\textsc{ResNet}\xspace}

\newcommand{\topTen}{\textsc{JavaTop10}\xspace}
\newcommand{\topFifty}{\textsc{JavaTop50}\xspace}

\renewcommand\fbox{\fcolorbox{black}{white}}

\newcommand{\approach}{\textsc{Code2Snapshot}\xspace}

\begin{document}

\title{\approach: Using Code Snapshots for Learning Representations of Source Code}


\author{
    \IEEEauthorblockN{Md Rafiqul Islam Rabin}
    \IEEEauthorblockA{
        \textit{mrabin@uh.edu} \\
        University of Houston\\
        Houston, TX, USA
    }
    \and
    \IEEEauthorblockN{Mohammad Amin Alipour}
    \IEEEauthorblockA{
        \textit{maalipou@central.uh.edu} \\
        University of Houston\\
        Houston, TX, USA
    }
}

\maketitle
\thispagestyle{firstpage}
\begin{abstract}
There are several approaches for encoding source code in the input vectors of neural models. These approaches attempt to include various syntactic and semantic features of input programs in their encoding. In this paper, we investigate \approach, a novel representation of the source code that is based on the snapshots of input programs. We evaluate several variations of this representation and compare its performance with state-of-the-art representations that utilize the rich syntactic and semantic features of input programs.

Our preliminary study on the utility of \approach in the code summarization and code classification tasks suggests that simple snapshots of input programs have comparable performance to state-of-the-art representations. Interestingly, obscuring input programs have insignificant impacts on the \approach performance, suggesting that, for some tasks, neural models may provide high performance by relying merely on the structure of input programs. 
\end{abstract}


\section{Introduction}
\label{sec:introduction}

Deep neural models are widely used to develop code intelligence systems \cite{allamanis2018survey,harman2020survey}. 
These models are powerful learning tools that provide a large hypothesis class and a large capacity for learning almost any arbitrary function. However, the representation of input programs as vectors can have a significant impact on the performance of the models.

Several representations for source code have been proposed in literature \cite{chen2019embeddings, sharma2021survey}.
These approaches try to encode the various syntactic and semantic features of input programs in the code embeddings. 
For example, \ctv \cite{alon2019code2vec} or \cts \cite{alon2019code2seq} encode paths in the abstract syntax trees of programs, \ggnn \cite{allamanis2018ggnn} encodes the data and control dependencies of programs, \ftv \cite{DeFreez} encodes paths in the extended call-graphs of programs, and so on.

The underpinning insight is that by encoding more information about the program in the feature vectors, the code intelligence model would be able to extract more patterns and insights, and are likely to perform better.
Unfortunately, due to the opacity of the neural models, it is unclear what features these models capture from the input data. Recent studies suggest that models heavily rely on variable names \cite{rabin2019tnpa, compton2020obfuscation,rabin2021generalizability} and a few tokens \cite{allamanis2015suggesting,rabin2020demystifying,suneja2020probing,rabin2021sivand,rabin2022perses} for their prediction, and they can memorize data points for achieving high performance \cite{allamanis2019duplication,rabin2022memorization}.

In this paper, we explore a new source code representation, \approach, in which, instead of using the traditional syntactic or semantic program features, we use program snapshots to represent any arbitrary code snippets.
Additionally, we study the \emph{redacted} snapshots of input programs that merely represent the structure of input programs. The main idea is that in real settings, seasoned programmers may get a hunch by merely looking at the structure of input programs.

We report the result of our preliminary study on the utility of \approach for code summarization \cite{allamanis2016summarization} and code classification \cite{bui2019sa} tasks. 
We compare its performance with the complex state-of-the-art representations, \ctv and \cts, that use richer syntactic and semantic features based on tokens and paths. 
Experiments on the most frequent method names and algorithm classes suggest that the simple snapshots of input programs can provide performance comparable to the complex neural code embeddings. However, the gap between the performance of \approach and \cts grows as we increase the size of dataset.

\smallskip
\noindent\textbf{Contributions.} We make the following contributions.
\begin{itemize}[noitemsep,topsep=0pt]
    \item We propose several source code embeddings based on the snapshots and structure of the input programs. 
    \item We provide a preliminary result on the evaluation of our proposed embeddings for code summarization and code classification tasks. 
\end{itemize}

\smallskip
\Part{Artifacts}.
The code and data of this study will be publicly available at \url{https://github.com/mdrafiqulrabin/Code2Snapshot}.

\section{Related Work}

Several source code embeddings have been proposed to encode source code as input vectors to neural networks \cite{chen2019embeddings}. For brevity and page limitation, here, we only mention the most relevant to our work.

\citet{allamanis2014conventions} introduce a framework that uses sequences of tokens to represent the source code.
\citet{alon2019code2vec} propose \ctv which is a fixed-length representation of source code based on the bags of paths in the abstract syntax tree.
\citet{allamanis2018ggnn} propose \ggnn that includes a richer feature space including control and data dependencies from the source code.
\citet{hellendoorn2018type} proposed an RNN-based model using sequence-to-sequence type annotations for type inference.

Some works have been done in the area of using images of input programs for overcoming feature extraction challenges.
\citet{dey2019socodecnn} propose an approach for converting programs to images, where they use the intermediate representation of programs to generate a visual representation of input programs.
\citet{keller2020visualization} apply different visual representations of code and transfer learning for code semantics learning.
\citet{bilgin2021code2image} also uses a colored image of syntax trees for representing input programs.

Unlike previous works, this work mainly focuses on the model's reliance on the structure of input programs for intelligent code analysis.

\section{Methodology}
\label{sec:approach}
\begin{figure*}
    \centering
    \includegraphics[width=0.98\linewidth]{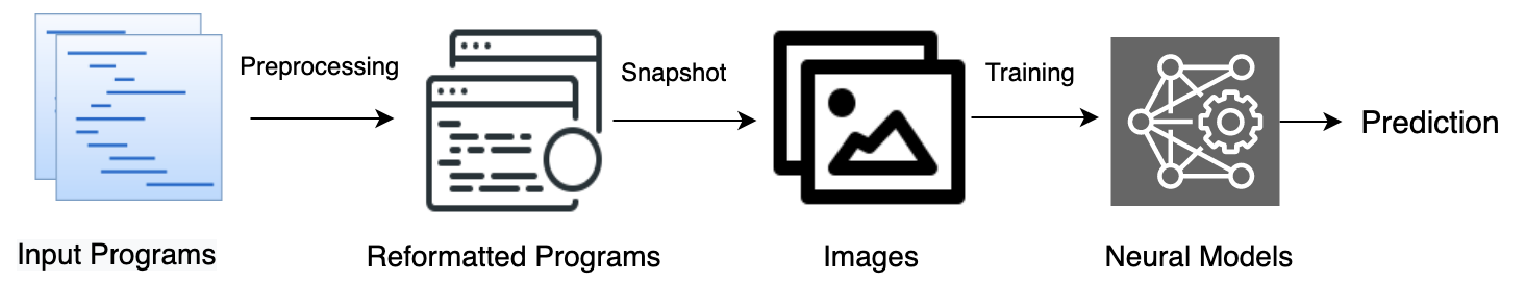}
    \caption{Workflow of the proposed \approach approach.}
    \label{fig:approach}
\end{figure*}

Figure~\ref{fig:approach} depicts a high-level view of our proposed approach.
Given the dataset of input programs, we reformat the original program and take snapshots from the reformatted program, and then train with Convolutional Neural Networks (CNNs) for predicting target labels.
Our approach contains the following key steps.

\subsection{Reformatting Code}
The original input program is often poorly aligned and has severe indentation problems such as extra spacing and newlines. It also includes random user comments and documentations. As a result, the original input program is not suitable for generating images, as we cannot get actual structural information from randomly aligned and commented code. To address these problems, we reformat the original program using \texttt{JavaParser} tool \cite{javaparser} that provides functions to analyze Java code. We remove documentations and comments from input programs, and adjust indentation of code by removing extra white spaces and line breaks.

\subsection{Taking Snapshots}
The Python Imaging Library (\texttt{PIL}) \cite{pil} provides image processing capabilities to add text on image.
After reformatting the original program, we use the \texttt{PIL} library to take snapshots from the reformatted program.
However, many programs contain more than a hundred lines of code and very long statements. Thus, the generated image from a larger program becomes extremely blurry when resized. For that reason, we shorten the size of a program by limiting statements up to $30$ lines and a maximum of $120$ characters per line. We simply filter out additional lines and characters from programs.
We follow the below steps to take snapshots.
\begin{itemize}
    \item Given a sample program, we first reformat the code and shorten the size of a larger program by filtering out extra lines and characters.
    \item Next, we create an image object with grayscale mode and white background and set the size of the image as a fixed window of $30$ lines and $120$ characters.
    \item We then adjust the spacing and font settings until finding a reasonable fit for the rendered text. In this work, we use `Times New Roman' as default font with $50$ points as requested size and $50$ pixels as line spacing.
    \item After that, we read the actual text from the reformatted program and write the text on the image in black color.
    \item Finally, we convert the image object to \texttt{PNG} format and save it to a directory.
\end{itemize}
Now, instead of any complex learning of source code embeddings, we can feed these images to CNN models and train for a downstream task.

\subsection{Training Models}
We train multiple CNN models on the snapshots of input programs for predicting target labels such as method name or algorithm class.
Before training, we apply image transformation and resize each image into ``\texttt{512 x 512}'' dimension in order to ensure that each image has the same size.
We also apply normalization on images to ensure that each pixel has a similar data distribution.
We train CNN models with the images of training set and validation set, and compute the performance on the images of test set.
During training CNN models, in the code summarization task, we use the image of a method body as input and the name of that method as output. Similarly, in the code classification task, we use the image of code snippets as input and the corresponding algorithm class as output.

\section{Experimental Settings}
\label{sec:settings}

\subsection{Task}
\label{sec:task}
We use two popular tasks to evaluate the \approach representation: (a) code summarization \cite{allamanis2016summarization}, such as method name prediction (\mnp), and (b) code classification \cite{bui2019sa}, such as algorithm classification (\cc). 
The \mnp is a supervised learning task wherein the inputs are the body of methods and the outputs are the corresponding method names.
For example, given the following code snippet: ``\texttt{void f(int a, int b) \{ int temp = a; a = b; b = temp; \}}'', a trained model aims to predict the method's name as ``\texttt{swap}''. 
Similarly, in the \cc task, a model learns to classify the code snippets into different algorithm classes, \eg, ``\texttt{bubble sort}''.

\subsection{Dataset}
\label{sec:dataset}

For \mnp task, we use \topTen and \topFifty, subsets of popular \JL and \JS datasets \cite{alon2019code2seq}, respectively.
The \topTen dataset contains the top $10$ most frequent methods from the \JL dataset. The training, validation, and test set contains a total of $10000$ ($1000$ instances per label), $4847$, and $7100$ examples, respectively.
The \topFifty dataset contains the top $50$ most frequent methods from the \JS dataset. The training, validation, and test set contains a total of $27119$, $1803$, and $4265$ examples, respectively. 

For \cc task, we use \SA dataset \cite{bui2019sa} that contains $1000$ sorting algorithms crafted from GitHub and labeled into $10$ algorithm classes. These are: bubble, bucket, heap, insertion, merge, quick, radix, selection, shell, and topological. The training, validation, and test set split is kept stratified at 70:10:20.

\begin{figure*}
    \centering
    \begin{subfigure}{.32\textwidth}
        \centering
        \caption{\textbf{\original}}
        \label{fig:img_original}
        \fbox{\includegraphics[width=0.98\linewidth,height=5cm]{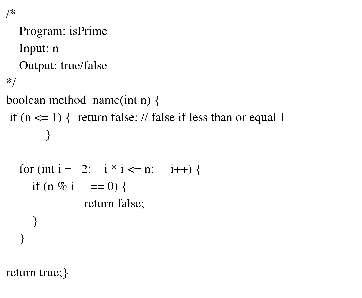}}
    \end{subfigure}%
    \begin{subfigure}{.32\textwidth}
        \centering
        \caption{\textbf{\refactor}}
        \label{fig:img_reformat}
        \fbox{\includegraphics[width=\linewidth,height=5cm]{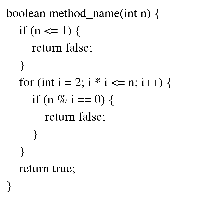}}
    \end{subfigure}%
    \begin{subfigure}{.32\textwidth}
        \centering
        \caption{\textbf{\xalpha}}
        \label{fig:img_xalpha}
        \fbox{\includegraphics[width=\linewidth,height=5cm]{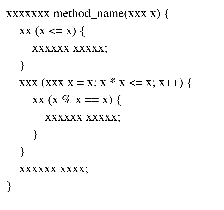}}
    \end{subfigure}
    \caption{Example of different code snapshots.}
    \label{fig:example_image_types}
\end{figure*}

\subsection{Input Types}
\label{sec:input}

\Part{Image.}
Using \texttt{PIL} \cite{pil}, we generate three variations of the \approach representation: a) \original, b) \refactor, and c) \xalpha.
\Cref{fig:example_image_types} depicts examples of these representations.
\begin{itemize}
    \item \original: We use the original input programs \emph{as is} to construct the visual representation of input programs. \Cref{fig:example_image_types}a shows an example of this representation where comments, extra lines and spaces remain unchanged.
    
    \item \refactor: We reformat the input programs to adjust indentation, remove comments. Moreover, for long programs, we include the first $30$ lines of code. \Cref{fig:example_image_types}b shows an example of the reformatted input program derived from the original input program of \Cref{fig:example_image_types}a.
    
    \item \xalpha: We replace any alphanumeric characters in the reformatted input programs with ``x". Punctuations and mathematical operators remain intact.
    For example, ``\texttt{int i = 2;}" becomes ``\texttt{xxx x = x;}". \Cref{fig:example_image_types}c is the example of \xalpha on the reformatted program of \Cref{fig:example_image_types}b.
\end{itemize}

\smallskip
\Part{Token.}
Using \texttt{JavaParser} \cite{javaparser}, we traverse the abstract syntax tree of program and collect all tokens from the method body. We then represent a program as a sequence of tokens. For example, \texttt{[int, i, =, 2, ;]} is the list of tokens for an expression ``\texttt{int i = 2;}''.

\smallskip
\Part{Path.}
Using \texttt{JavaExtractor} \cite{alon2019code2vec, alon2019code2seq}, we extract and preprocess a bag of path contexts from Java files. For example, a path context for the expression ``\texttt{int i = 2;}'' is ``\texttt{i, <NameExpr ↑ AssignExpr ↓ IntegerLiteralExpr>, 2}''.

\subsection{Models}
\label{sec:model}

\Part{Image-based Models.}
We use two classic convolutional networks: \alexNet (an 8-layers deep convolutional neural networks \cite{krizhevsky2012alexnet} that consists of five convolutional layers, then two fully-connected hidden layers, and one fully-connected output layer) and \resNet (an 18-layer residual networks \cite{he2016resnet} that consists of one convolutional input layer, then four modules made up of two residual blocks with two convolution layers in each block, and a final fully connected layer).


\smallskip
\Part{Token-based Models.}
We create a simple recurrent neural network as baseline for learning method names from the sequence of tokens \cite{allamanis2014conventions,hellendoorn2018type} in method body. The network includes an embedding layer followed by a 2-layers bi-directional LSTMs and a final fully connected linear layer with softmax activation function. 
In the embedding layer, we use EmbeddingBag which computes the embedding of a method by taking the average of all token embeddings.

\smallskip
\Part{Path-based Models.}
We use two popular path-based models in this study: \ctv \cite{alon2019code2vec} and \cts \cite{alon2019code2seq}. The models use a bag of paths between two terminal nodes in the abstract syntax tree (AST) for representing an arbitrary source code as an embedding vectors. While \ctv considered an entire path as single entry and learns monolithic path embeddings, \cts sub-tokenized each path and uses LSTMs to encode paths node-by-node.



\smallskip
\Part{Training Models.}
For each combination of dataset, model, and representation, we train a model up to $100$ epochs. For training image-based models and token-based models, we use stochastic gradient descent optimizer and cross-entropy loss function. We train both \alexNet and \resNet with the original configurations but modified the input and output layers for adjusting the size of input images and the number of target classes, respectively.
For path-based models, we train both \ctv and \cts with the configurations used in the original work with the batch size of $128$. Following the evaluation metrics commonly used in the literature \cite{alon2019code2vec, alon2019code2seq}, we use the accuracy, precision, recall, and F1-score as  metrics.

\section{Results}
\label{sec:results}
\begin{table}
    \centering
    \caption{Results for different models and code embeddings.}
    \vspace{1mm}
    \label{table:result_images}
    \def\arraystretch{1.5}
    
    \subcaption{Task: \mnps}
    
    \resizebox{0.99\columnwidth}{!}{%
    \begin{tabular}{|c|c|c|c|c|c|c|} 
        \hline
        Dataset & Model & Input Types & Precision & Recall & F1-Score & Accuracy \\ \hline \hline
        
        \multirow{9}{*}{\topTen}
        & \multirow{3}{*}{\alexNet} & \original & 55.25 & 52.49 & 53.43 & 52.49 \\ \cline{3-7}
        &  & \refactor & 82.84 & 81.18 & 81.26 & 81.18 \\ \cline{3-7}
        &  & \xalpha & 80.26 & 78.58 & 78.68 & 78.58 \\ \cline{2-7}
        & \multirow{3}{*}{{\resNet}} & \original & 65.03 & 61.76 & 61.83 & 61.76 \\ \cline{3-7}
        &  & {\refactor} & \underline{85.10} & \underline{83.80} & \underline{83.78} & \underline{83.80} \\ \cline{3-7}
        &  & \xalpha & 83.75 & 83.06 & 83.00 & 83.06 \\ \cline{2-7}
        & \bilstm & \ttoken & 69.72 & 67.87 & 67.52 & 67.87 \\ \cline{2-7}
        & \ctv & \multirow{2}{*}{{\tpath}} & 85.80 & 84.96 & 84.85 & 84.96 \\ \cline{2-2}\cline{4-7}
        & \cts &  & \textbf{86.88} & \textbf{86.65} & \textbf{86.50} & \textbf{86.65} \\ \cline{1-7}
        
        \multirow{9}{*}{\topFifty}
        & \multirow{3}{*}{\alexNet} & \original & 33.19 & 26.10 & 27.63 & 26.10 \\ \cline{3-7}
        &  & \refactor & 56.88 & 51.72 & 51.30 & 51.72 \\ \cline{3-7}
        &  & \xalpha & 55.67 & 51.09 & 50.75 & 51.09 \\ \cline{2-7}
        & \multirow{3}{*}{{\resNet}} & \original & 58.09 & 45.25 & 48.96 & 45.25 \\ \cline{3-7}
        &  & {\refactor} & \underline{62.38} & \underline{62.88} & \underline{61.44} & \underline{62.88} \\ \cline{3-7}
        &  & \xalpha & 62.56 & 59.60 & 58.52 & 59.60 \\ \cline{2-7}
        & \bilstm & \ttoken & 48.72 & 44.67 & 41.82 & 44.67 \\ \cline{2-7}
        & \ctv & \multirow{2}{*}{{\tpath}} & 69.12 & 68.11 & 67.39 & 68.11 \\ \cline{2-2}\cline{4-7}
        & {\cts} &  & \textbf{73.73} & \textbf{71.40} & \textbf{71.59} & \textbf{71.40} \\ \hline
    \end{tabular}%
    }
    
    \bigskip
    \subcaption{Task: \ccs}
    
    \resizebox{0.99\columnwidth}{!}{%
    \begin{tabular}{|c|c|c|c|c|c|c|} 
        \hline
        Dataset & Model & Input Types & Precision & Recall & F1-Score & Accuracy \\ \hline \hline
        \multirow{5}{*}{\SA}
        & \multirow{3}{*}{{\resNet}} & \original & 40.96 & 33.33 & 29.00 & 33.33 \\ \cline{3-7}
        &  & {\refactor} & 44.87 & 46.30 & 44.64 & 46.30 \\ \cline{3-7}
        &  & \xalpha & \underline{76.78} & \underline{75.46} & \underline{75.58} & \underline{75.46} \\ \cline{2-7}
        & {\ctv} & \multirow{2}{*}{{\tpath}} & \textbf{87.94} & \textbf{87.85} & \textbf{87.81} & \textbf{87.85} \\ \cline{2-2}\cline{4-7}
        & \cts & & 77.91 & 77.10 & 76.39 & 77.10 \\ \hline
    \end{tabular}%
    }
    
\end{table}

\Cref{table:result_images} shows the performance of different models and embeddings across various tasks and datasets of source code. The \underline{underlined} values denote the best result in \approach variations and the values in \textbf{bold} denote the best code embeddings overall.

\subsection{Performance of the \approach Variations}

In this section, we compare the effectiveness of different variations of \approach representation on the \mnp and \cc tasks.

According to the results in \Cref{table:result_images}, 
\refactor has consistently performed much better than \original, more than 20 percentage points.
Training CNN models with the images of original program perform poorly, perhaps because the original input programs often suffer from inconsistent indentation, random comments, large program body, etc. After reformatting the input programs and training with the images of reformatted programs, each CNN model significantly increases its performance in predicting method names or classifying code snippets into algorithms.

On average, in \mnp task, the performance of the \resNet and \alexNet models is between $20$ and $30$ percentage points higher on \refactor images compared to \original images. 
Comparing among CNN models, we can see that \resNet model achieves higher accuracy than \alexNet models in both \topTen and \topFifty datasets. 
The \resNet model on \refactor images achieve more than $80\%$ accuracy in \topTen dataset, and more than $60\%$ accuracy in \topFifty dataset.

Next, in \cc task, we train the best \resNet model with \SA dataset and observe that the \resNet model achieves more than $45\%$ accuracy on \refactor images which is around $13\%$ higher than \original images.

The overall result seems consistent across the tasks and datasets used in this experiment. It may suggest that the visual code embeddings have the potential to be used for the tasks like \mnp and \cc.

\observation{We get significant performance improvements with the image of refactored programs over original programs across all datasets and CNN models.}

\subsection{Comparison with State-of-the-art Code Embeddings}
Here, we compare \approach with the baseline token-based embeddings and the state-of-the-art path-based embeddings.

For \mnp task, \Cref{table:result_images} shows that the \alexNet model of \refactor or \xalpha images consistently outperforms the \bilstm model of tokens by more than $10$ and $5$ percentage points in \topTen and \topFifty datasets, respectively. On the same setting, the \resNet model outperforms the \bilstm model by around $15$ percentage points in both datasets.

Comparing \approach with path-based embeddings, we can see that, in the \topTen dataset, the \resNet model of \refactor images performs very close to the \ctv model, and the \cts model's accuracy is only $3$ percentage points higher.
In the \topFifty dataset, the gap between \approach models and path-based models grows, as the accuracy of \ctv and \cts models are $5$-$8$ percentage points higher than the best accuracy of \resNet models. On the other hand, in \SA dataset, the \resNet model of \xalpha images performs very close to the \cts model but has more than $10\%$ gap with the \ctv model.

\observation{\refactor significantly outperforms token-based embeddings in both \topTen and \topFifty datasets. Path-based embeddings perform slightly better than \refactor and \xalpha.}

\subsection{Impact of Obscuring Images}
To get an intuition of what actually the \approach models learn, 
we use \xalpha representation where we replace any alphanumeric character with ``x'' (\Cref{fig:example_image_types}c). 
In our experiments, \xalpha performs almost the same as \refactor in \mnp task and around $30$ percentage points higher in \cc task.
The results of \xalpha may suggest that models highly learn from the structure of programs. For example, for \mnp and \cc tasks, models can predict the method names or classify the code snippets merely by identifying the structure of input programs, respectively.

\observation{\xalpha performed very close to (or even better than) \refactor, suggesting that perhaps neural models merely capture the structure of input programs for some tasks.}

\section{Discussion and Future Work}

Neural models are opaque. It is hard to pinpoint what patterns do they learn from the input data and what is the best representation of the inputs to optimize their performance. 
We believe that our results show promising directions to better understanding the neural code intelligence models in general, and the representation of the source code in particular.
The results of \xalpha may suggest that the neural models merely rely on the structure of the code. The encoding uses only black and white images for representing the source code. In the future, we plan to investigate if we enrich the feature vectors by colors, \ie, choosing different colors for different program constructs, or choosing the same colors for the same variable names, we can gain meaningful improvements. The main insight is that it might help the neural models to extract more complex patterns, in addition to the structure of the code.

Moreover, a potential explanation for the comparable performance of \approach in \mnp and \cc tasks might be that, in the real setting, the visual structure of code can guide developers to guess the labels. For example, a one line method is probably a getter-setter method. Therefore, it is unclear if \approach can provide similarly promising results for tasks that require reasoning about the input programs, \eg, variable misuse or defect detection. We plan to explore the effectiveness of \approach in such tasks. 
Despite our best effort, it is possible that experiments with different models, sizes of images, and font settings may produce different results. Our further plan also includes a detailed study with a variety of code intelligence tasks, embeddings, and datasets.

Lastly, the computer vision research community has made great advances in the interpretability of neural models \cite{zeiler2014visualizing,selvaraju2017grad}. 
Representing source code as an image and using convolutional neural architectures would enable using the most recent techniques devised by that community and translating the results to the field of code intelligence.

\section{Conclusion}
In this paper, we investigated \approach that encoded the source code as a simple black and white image and provided a preliminary evaluation of its effectiveness in \mnp and \cc tasks. 
We observed that this simple visual encoding is surprisingly effective in our study, even when replacing all alphanumeric characters of programs with a random letter `x'. The results warrant further investigation on the applicability of \approach for other tasks and datasets. These initial results can enable further research to answer fundamental questions about the neural code intelligence models, \ie, what do they actually learn? and how they can be improved? 
We also plan to explore in such directions.

\balance
\bibliography{refs}
\bibliographystyle{IEEEtranN}

\end{document}